\documentclass[twocolumn,floatfix,pre,showpacs]{revtex4}
\usepackage{graphicx}
\usepackage{amssymb}
\begin{document}

\title{Simple models for granular force networks}
\author{John F. Wambaugh\footnote{Current address: National Center for Computational Toxicology, US EPA, Research Triangle
Park, NC 27711}}
\affiliation{
Departments of Physics and Computer Science and Center for Nonlinear and
Complex Systems Duke University, Durham, NC 27708}
\email{wambaugh@phy.duke.edu}

\begin{abstract}
A remarkable feature of static granular matter is the
distribution of force along intricate networks.  Even regular
inter-particle contact networks produce wildly inhomogeneous force
networks where certain ``chains" of particles carry forces far larger than
the mean.  In this paper, we briefly review past theoretical approaches to understanding the geometry of force networks.  We then investigate the structure of experimentally-obtained granular force networks using a simple algorithm to obtain corresponding graphs.  We compare our observations with the results of geometric models, including random bond percolation, which show similar spatial distributions without enforcing vector force balance.  Our findings suggest that some aspects of the mean
geometry of granular force networks may be captured by these simple descriptions.
\end{abstract}


\maketitle

\section{Introduction}

Dense granular materials are composed of many particles interacting through multiple, persistent contacts.  Without knowing the history of a particular granular assembly, the mobilization of the friction of each contact is indeterminate \cite{Halsey99}.  Even without friction, the simplest granular materials, such as packings of macroscopically uniform spheres, can still display highly non-uniform networks of contacts --- any difference in the size of particles, including microscopic aberrations on the surface of ``smooth" spheres, is sufficient to create a complicated contact network \cite{Roux89}.

\begin{figure}
\centering
\includegraphics[width=2in]{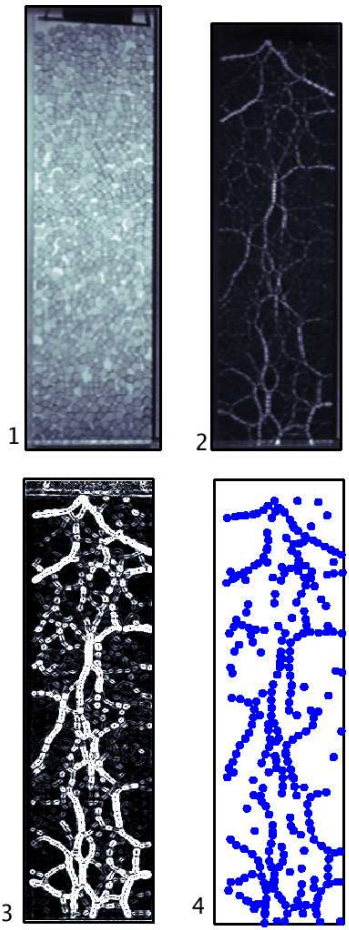}
\caption[Depiction of algorithm for determining force network]{Analysis of the two-dimensional granular assembly of bi-disperse polymer disks depicted in {\bf 1}.  The contact network for the disks can be thought as a series of vertices centered at each disk connected by edges wherever two disks are in contact.  {\bf 2} Because the disks are
photoelastic, when they are illuminated between crossed polarizers bright fringes appear in disks under stress allowing force
chains to be clearly seen.  A load has been applied to the top of the pile
causing the creation of additional contacts and force chains supporting
the load.  {\bf 3} We examine the gradient of the stress image as a rough measure of the number of fringes in each disk, which is proportional to the applied force.  {\bf 4} Finally, we apply an algorithm to the gradient image to find a graph describing the force network.}
\label{silo}
\end{figure}
Experiments show that for any given particle within a granular material, the fluctuations in the magnitude of the force carried by that particle can be extremely large \cite{Dantu68,Baxter97}.  This is the result of long-range force
correlations that can be characterized as a {\em force network} in
which some bonds of the contact network carry forces much larger than the
mean force on all contacts.  As in Fig.~\ref{silo}.2, forces propagate along roughly linear ``chains" of contacts.  Because the
force network depends upon the contact network, the propagation direction
and distribution of the chains depends strongly upon the history of the
granular material \cite{Clement01a,Radjai97}.

The overall distribution of forces in a force network is roughly exponential, indicating the possibility for extremely large forces.  Since tangential forces are supported only by friction, their distribution falls monotonically, while forces along the chains are peaked at a mean value before falling exponentially \cite{Behringer05}.
As grains are added, if the force network cannot support the granular material, the material will rearrange until the new contact and force
network supports the granular material.  For this reason it has been suggested that granular materials exhibit self-organized criticality, in which the force
network behaves as an order parameter in response to displacements of the
material \cite{Roux89}.

For any configuration of grains (e.g., Fig.~\ref{silo}.1), if the
position and shape of each grain is known then a contact network can be generated describing the contacts between the grains.  The equations describing a granular material can be grouped into the known equations describing the behavior of the individual grains and the unknown equations describing the contacts between the grains as a function of material assembly and history. These unknown equations can be
cast as undetermined relationship between inter-grain forces and the
geometry of the contact network \cite{Edwards01}.

Attempts to resolve the indeterminacy of the constitutive relations have
focused upon statistical approaches similar to thermodynamics.  Ideally,
it might be possible to determine aggregate properties of granular
materials that do not depend upon knowing the specifics of the grains to
arbitrary precision.  Unfortunately there are two problems that have made
such an approach non-trivial.  First, a granular material at rest is only
in a quasi-equilibrium.  An arbitrarily small addition of energy can cause
irreversible (plastic) rearrangement of the material.  This is in part due
to the second problem, the lack of an analog to temperature to
``thermalize" the system.  The actual energy due to temperature is far too
small to move grains and, in the static case, there is no other quantity allowing the material
to explore phase-space for lower energy states.

Despite these complications, there is a rich tradition of models to understand both the distribution of magnitude and the spatial correlation of forces within granular matter \cite{Coppersmith96,Socolar97,Wittmer98}.  The applicability of these models to actual granular materials has been limited in part by being restricted to regular lattices of grains.  Recent numerical work \cite{Nienhuis05,Nienhuis06,Nienhuis06b}, however, suggests that the spatial distribution of force within irregular arrangements of grains might indeed be characterized through comparison with models with regular geometries.

Typically the force network is determined using a
threshold --- for instance, only bonds carrying forces larger than some value, often the
mean force, are considered to be part of the force network.  In numerical simulations, critical-like transitions from a sparsely distributed network of force chains to a percolating network have been observed as the threshold criteria for including bonds is varied.  In one case, the magnitude of the inter-grain mobilization of friction \cite{Matsuoka01b} was used and in another that magnitude of inter-grain forces \cite{Nienhuis06}.  It is important to note that these thresholded force networks will not be rigid, since they lack bonds that, while carrying only small amounts of force, provide rigidity to the network.

\begin{figure}
\centering
\includegraphics[width=1in,clip=]{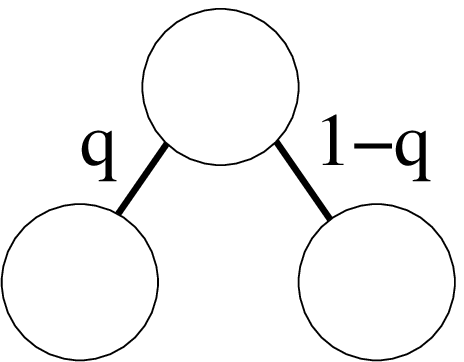}
\includegraphics[width=1.3in,clip=]{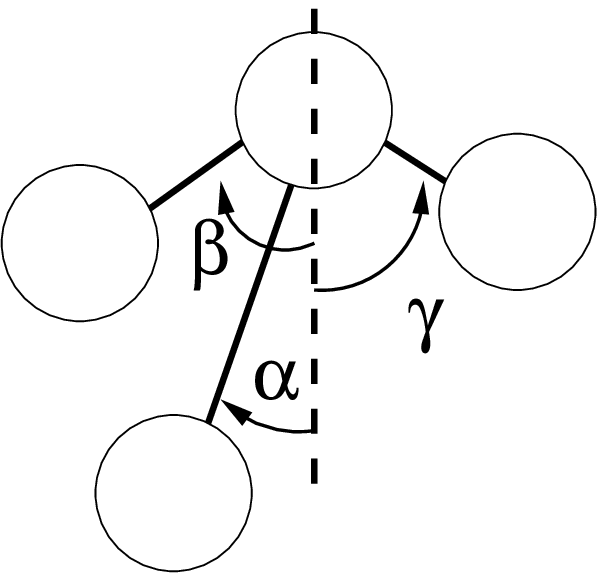}
\caption[Depiction of the $q$-model and the geometric version]{{\bf Left}
In the two-dimensional $q$-Model, a fraction $q$ of the total weight on
a particle on a regular lattice is transferred to one lower neighbor, and
the remaining $1-q$
is transferred to the other.  As originally formulated by Coppersmith, et
al., the values of $q$ can be drawn randomly from different
distributions.  In the ``critical" case, the value of $q$ is limited to
either $0$ or $1$.  {\bf Right} To make rough analogy to the $q$-Model
for more complicated geometries,
we use a geometry-based $q$-Model in which the fraction of a
vertex's weight that is
distributed along a downward pointing edge is proportional to the sine
of the angle that edge forms with the vertex.}
\label{geometricqmodel}
\end{figure}

There have been many attempts to model the behavior of force networks,
most famously the ``$q$-Model" of Coppersmith (1996).  In the
two-dimensional $q$-Model, depicted in Fig.~\ref{geometricqmodel},
granular materials are represented as a lattice
of sites with two upward neighbors and two downward neighbors.  Each site
has an intrinsic weight that is added to any weight it receives from the
neighbors above it.  The total weight on a site is distributed between the
two downward neighbors with one receiving a fraction $q$ of the total
weight and the other receiving $1-q$.  The $q$-Model allows the
determination of the distribution of weight by working downwards from the
top of the lattice.  This process can be shown to be diffusive-like in the
continuum limit \cite{Coppersmith96}.

Diffusive systems, however, do not allow for long-range propagating forces like those seen in force networks.  Experiments on small systems have shown that even averaging over multiple force chain networks does not support a diffusive description of granular materials \cite{Behringer03}.  However, in the case where $q$ is allowed to only take the values of $1$ or $0$, the $q$-Model does show chain-like clusters of bonds.  This case will be referred to as the ``critical $q$-Model" \cite{Coppersmith96}.

In all cases the $q$-model is a {\em scalar} model in that the distribution of weight, not force, is being described.  The forces exerted upon a particular lattice site do not need to balance, and this in part explains the lack of chain-like structures in all but the critical case.  {\em Vector} models that attempt to include force balance also exist.  One of the simplest lattice models that includes force balance is the ``tripod" model that, in two-dimensions, assigns each site three downward neighbors, one directly below and two offset to either side by a fixed angle.  In the continuum limit this model can give wave-like equations for propagating stress \cite{Wittmer98}.  Additionally, other models have included torque-balance on each site, giving meaning to the finite extent of the grains \cite{Socolar97}.  Including both force and torque balance allows derivation in simple cases for equations describing force equilibrium as functions of the contact distribution \cite{Edwards99}.
\begin{figure}
\centering
\includegraphics[width=3.3in]{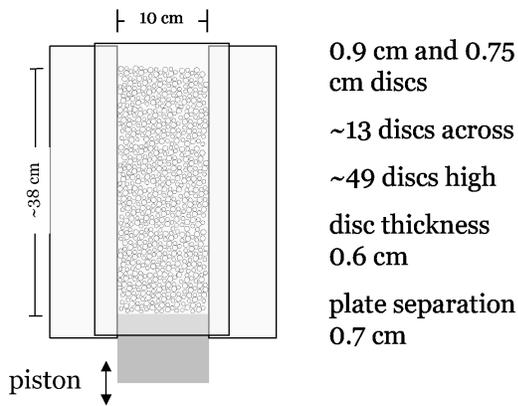}
\caption{In our experimental silo for observing granular force networks bi-disperse disks made of a photoelastic polymer are confined vertically to two-dimensions between clear plastic sheets and aluminum spacers serving as sidewalls.  The grains rest upon
a piston that is lowered slightly to make uniform the mobilization of friction at the walls.  The entire silo is illuminated between crossed polarizers, which allows fringes induced by applied stress on each grain to be imaged with a CCD camera.}
\label{silosetup}
\end{figure}

\section{Geometric Approaches to Granular Matter}

Research into the general behavior of force networks in granular matter has shown that certain properties, including the exponential distribution of large forces, should be expected to arise regardless of the grain-scale model.  In this case, some properties of granular materials may be phenomena of the statistics of force networks, and not the micro-mechanics of the grains \cite{Bagi97,Goddard04}.

This insight has led to approaches to granular materials from the mesoscopic level.  By assuming the presence of force chains and describing their behavior, it may be possible to avoid some of the complexity of granular materials and still achieve meaningful results \cite{Socolar02,Snoeijer04}.  In particular, network approaches to granular materials may bear fruit because the statistical mechanics of networks in general are beginning to be understood \cite{Newman04}.

The critical $q$-model is similar to random bond percolation on a lattice.
Percolation theory provides a framework for understanding how small
clusters of connected sites aggregate into larger clusters as a function
of the probability that a bond is selected. There are well known values,
depending upon the geometry of the lattice, for transitions from loose
aggregates of clusters to system-spanning, ``infinite" clusters
\cite{Sykes64}. In the case of the critical $q$-model, bonds are selected
at random from a set of edges representing the connections between the
lattice sites with the restriction that each lattice site has only one
load-bearing downward bond.  The set of load-bearing bonds can be thought
of as a subset of edges that, with the vertices as the lattice sites,
forms a sub-graph of the contact network.  Since weight at a given vertex is equal to one plus the size of the tree rooted at that vertex, the distribution of cluster sizes
gives bounds on the distribution of weights
%
%

In disordered granular materials the network of inter-grain contacts may perhaps
 serve in place of a fixed lattice for percolation.
Taking the contact network to be a graph, then the force network may be thought of as the sub-graph consisting of edges with greater than some threshold force.

Properly selecting the edges that comprise the force network requires consideration of vector force balance for each particular network.
  However, recent work has suggested that granular force networks belong to a specific universality class that can be characterized through analogy to random bond percolation, allowing the distributions of forces in a wide range of isotropic and anisotropic granular systems to be scaled onto a single distribution \cite{Nienhuis05, Nienhuis06, Nienhuis06b}.

 In this paper we make experimental observations of the force networks within a granular system and use a simple approach to extract the statistical distributions of graph properties for these force networks.  We characterize force networks both in terms of the graph-theoretic properties of degree distribution and cluster-size, as well as $q$-model inspired properties, including the generalized $q$-model weight and the size of the cluster resting on a given grain.  We then compare our experimentally-observed graphs with those generated by simple geometric models to determine in what ways simple statistical approaches do and do not describe the magnitude and spatial distribution of forces within dense granular matter.

\section{Methodology}

We use the experimental setup depicted in Fig.~\ref{silosetup} to observe
force networks within actual granular matter.  We confine 650 bi-disperse (one quarter
diameter $0.9$ cm and the rest $0.75$ cm) photoelastic polymer disks within a vertical, two-dimensional silo.  When illuminated between crossed polarizers, photoelastic materials show bright fringes of light in proportion to the stress applied to the material.  By taking the gradient of an image of a photoelastic disk along multiple directions we effectively ``count" the number of fringes and, using an empirical, linear calibration, can obtain the force distribution throughout the granular
material (see Fig.~\ref{silo}.3).

Using the gradient image of a particle force network we convolve with an image representing a single stressed grain to locate the positions of all stressed grains in the image.  We then assign an edge between all vertices separated by a distance in the range of $l/4$ to $l$ where $l$ is a manually-adjusted parameter.  The lower range is necessary because some fringe patterns within a single disk produce multiple points
when convolved, which has the effect of artificially increases clustering
of the graph.

Adding an edge to any grain that already has the threshold value $k_{max}$
edges causes the longest edge to that vertex to be removed.  We use the
smallest $k_{max}$ that does not significantly alter $P(k)$.

The specific values of the network-finding parameters used depends upon
many properties of the images of force networks, including the
image resolution and dispersion of the disks.  These parameters, in
addition to the correlation threshold for locating vertices, must be tuned to
produce force network graphs that accurately describe force networks.
We assigned edges within the range $l = 13$ pixels and we allowed a maximum of $k_{max} = 5$ edges.  As discussed below, the distributions reported here do not change unexpectedly with variation of correlation threshold.  The set of vertices and edges found this way are the experimentally obtained force network (see Fig.~\ref{silo}.4).

Using ensembles of force networks we can determine the probability distribution function for the average number of contacts per vertex and number distribution of cluster sizes.  We can also record the observed force at each vertex and create an ensemble probability distribution for force.

\begin{figure}
\centering
\includegraphics[height=3.3in,angle=270,clip=]{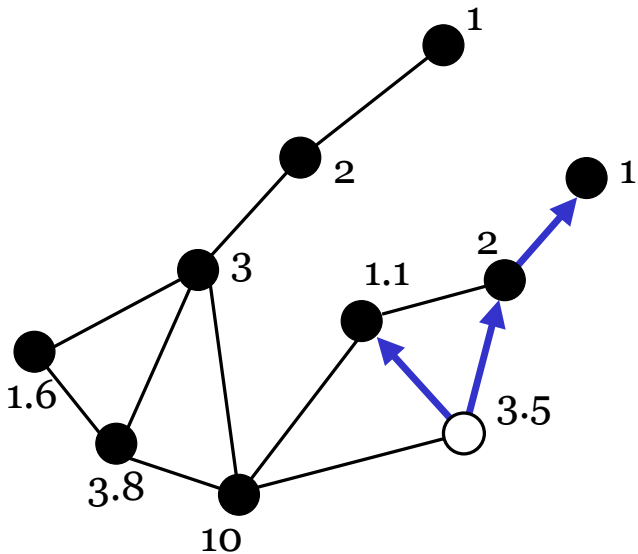}
\caption[Illustration of graph quantities measured]
{Illustration of the graph quantities measured.  The cluster size
$s = 10$ includes all the vertices.  The size of the upward tree rooted at
the white circle (indicated by directed edges), $t = 4$ includes all
vertices connected by upward edges only.  The degree of the white node is
$k = 3$.  The numbers indicate approximate geometric $q$-Model weights
$w$, determined for each vertex by assigning each an inherent unit weight
and distributing the weight among all downward edges in proportion to the
sine of the angle each edge makes with the vertex.}
\label{graphquantities}
\end{figure}
Figure~\ref{graphquantities} illustrates several graph quantities that we
measure.  In addition to cluster size and vertex
degree, we measure the size of the upward tree rooted at each vertex.
The upward tree is a graph made by treating each edge as a directed edge
oriented against gravity.  The size of the upward tree corresponds to the
number of vertices that are at least partially supported by a given vertex.
As illustrated in Fig.~\ref{geometricqmodel}, weights can be assigned to
each vertex and propagated along each downward pointing edge in
proportion to the cosine of the angle each edge makes with the vertex.
This allows for a $q$-Model-like weight distribution to be generated for  non-lattice geometries.  For all of these quantities an aggregate probability density function is
calculated.

\section{Results}

\subsection{Experimental Results}

\begin{figure}
\centering
\includegraphics[width=3.3in]{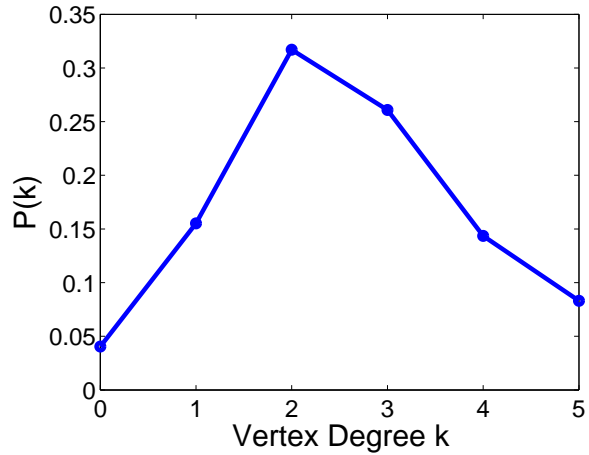}
\caption[Distribution of vertex degree for experimental force networks]{The distribution of vertex degree peaks at $k=2$.  The lack of a sharp uptick at $k=5$ indicates that truncating at 5 has not significantly affected the distribution.    These results are from analyzing one hundred experimentally-obtained force networks under 56g load.}
\label{forcedatak}
\end{figure}
We analyzed one hundred force networks obtained from experimental images of grains in a vertical silo subject to a $56$ g overload that created an approximately uniform distribution of force chains, as in Fig.~\ref{silo}.2.  Figure~\ref{forcedatak} shows the distribution of vertex degrees and, as expected for chain-like structures, vertices most often have two edges. From geometric considerations, the maximum number of contacts for a single disk in a bi-disperse mixture of disks with radii $r_{small}$ and $r_{large}$ is:
\begin{equation}
k_{max} = \pi/2/\arcsin{\left[r_{small}/2/\left(r_{small}+r_{large}\right)\right]}\nonumber
\end{equation}

In this case $k_{max} = 6.85$ and there is no probability of seeing a vertex with degree higher than six.  Thus, as Fig.~\ref{forcedatak} shows, the imposed cutoff of $k_{max} = 5$ used in the edge finding routine does not significantly affect the distribution of $P(k)$ since fewer than a tenth of all vertices have degree larger than $4$.

\begin{figure}
\centering
\includegraphics[width=3.3in]{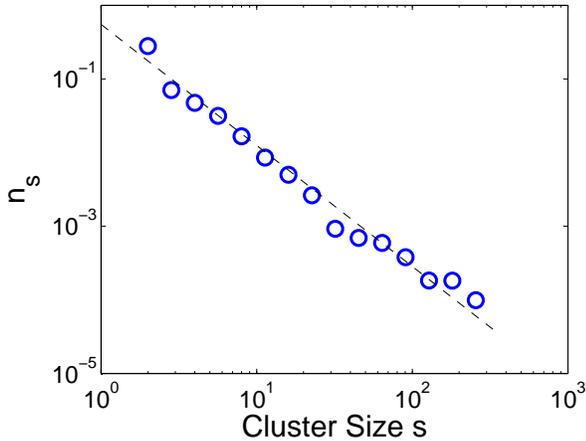}
\caption[Distribution of cluster sizes for experimental force networks]{The distribution of $n_s$, the number of clusters of size $s$ divided by the total number of clusters, appears to be roughly linear over many decades of probability and two decades of cluster-size (corresponding to the system size) when binned logarithmically.  A power-law distribution would support the idea of Roux et al. \cite{Roux89} that the force network may perhaps be considered a critical phenomenon.}
\label{forcedatas}
\end{figure}
The distribution of the number of observed clusters
of size $s$ (as a fraction of the total number of clusters) appears to follow a power law, as indicated by Fig.~\ref{forcedatas}.  This behavior lends support to the suggestion of Roux, et al. \cite{Roux89} that the force network may be an example of a self-organized critical phenomena.  Fits were obtained for regressions to the data (neglecting the end points) for a power law ($n_s = a_1 s^{a_2}$), the product of a power law and an exponential ($n_s = b_1 s^{b_2} e^{b3 s}$), and a log-normal ($n_s = \frac{c1}{s}e^{c_2\left(s - c_3\right)^2}$) distribution.  In all cases, the ratio of the variance predicted by the regression to the observed variance ($R^2$) is greater than  $0.95$, indicating likely fits \cite{Moore03}.  The fits for the power law-exponential and log-normal distributions, however, both produce positive curvature indicating increasing likelihood with very large cluster size.  We believe such distributions are  unphysical since increasing probability with increasing cluster size is not normalizable, so that a power law with an exponent $a_2 = -1.6$ seems most likely.

\begin{figure}
\centering
\includegraphics[width=3.3in]{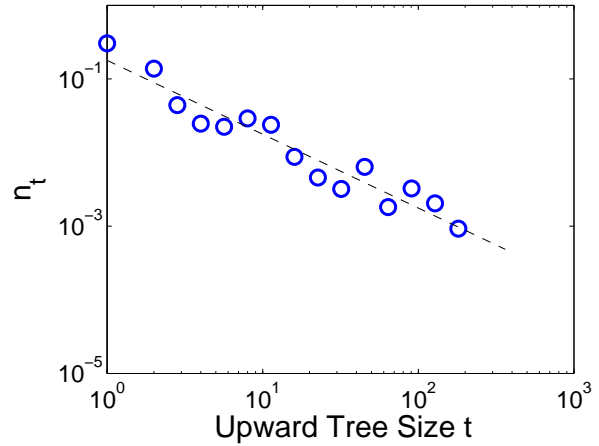}
\caption[Distribution of upward cluster sizes for experimental force networks]{The logarithmically-binned distribution of $n_t$, the number of upward clusters of size $t$ divided by the total number of such clusters, appears linear on a log-log plot, similar to the cluster distribution in Fig.~\ref{forcedatas} but with a shallower slope.}
\label{forcedatat}
\end{figure}
As shown in Fig.~\ref{forcedatat}, the number distribution of upward clusters is shallower ($a_2 = -1.0$) than for whole clusters but still appears to be linear.  We expect the distribution of upward trees to generally be shallower than the distribution of clusters since each large cluster contains many upward trees.  Regressions to power law, power law-exponential, and log-normal distributions all have $R^2 > 0.85$, but only the fits for the power law-exponential distribution give the observed drop-off with large upward tree size.

\begin{figure}
\centering
\includegraphics[width=3.3in]{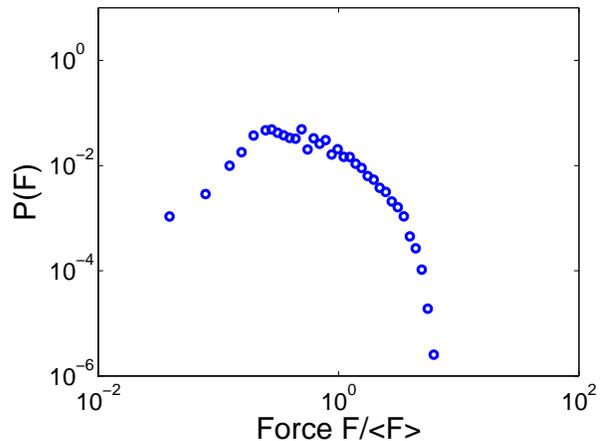}
\caption[Distribution of forces in experimental force networks]{The force observed at each vertex computed from the gradient is non-monotonic and the greatest probability is for forces below the mean force in the network (indicated by $F/\langle F\rangle = 10^0$).}
\label{forcedataF}
\end{figure}
Figure \ref{forcedataF} shows the probability distribution of intensity of the gradient squared image at the pixel location of each vertex.  The intensity corresponds with the force on the disk at that vertex and thus Fig.~\ref{forcedataF} is an approximate force probability distribution.  This non-monotonic distribution peaks below the mean force and then falls off faster than a power law.

\begin{figure}
\centering
\includegraphics[width=3.3in]{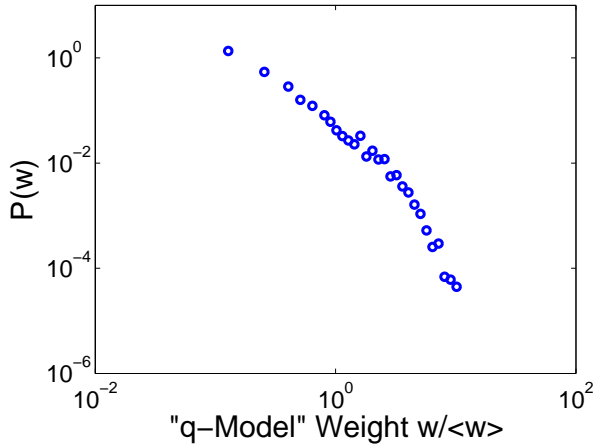}
\caption[Distribution of geometric $q$-Model weights assigned to experimental force networks]{The probability density function of weight assigned to vertices in the experimentally-obtained force networks by a method similar to the q-Model.  Unlike the actual force data, the curve is monotonic except for a slight dip.}
\label{forcedatamodels}
\end{figure}
We then compare the actual force distribution to what would be predicted using our geometric $q$-Model, obtaining the distribution in Fig.~\ref{forcedatamodels}.  The $q$-Model
weights are sub-power law but nearly monotonic, unlike the actual force data.

The geometric limit of $6$ possible contacts per disk corresponds to $6$
potential edges per vertex, however the average number of contacts
(coordination number) for a disordered granular material is $4$
\cite{Roux89}.  Assuming that there are four contacts per vertex, we can
calculate a chance $p$ that a bond carrying a large force is present.
For a particular vertex $i$, $p_i = k_i/4$ and the average value $p \equiv
\sum_{i=1}^N{p_i/650} = \sum_{i=1}^N{k_i/4/650}$, where $N$ is the total
number of vertices in the force network graph, and 650 is the total number
of particles.  For the observed networks, $p = 0.2214$.

Additionally, since we expect $4$ contacts on average, Fig.~\ref{forcedatak} indicates that grains in the force network are typically in contact with one or two grains not in the force network.

For each vertex we can also compute the clustering coefficient --- how many
edges are shared by the neighboring vertices $4\choose 2$ possible edges \cite{Watts98}.  Unlike a random network where any vertex may be contact with an unlimited number of other vertices at any position, the choice of our radii and the quasi-two-dimensional geometry of our silo restricts the number of neighbor-neighbor contacts between $n$ neighbors to $n-1$.  We calculate $c = 0.1219$ for the observed force network.  This amount of clustering is similar to that seen in scale-free networks that exhibit power law behavior, such as the World-Wide Web and power grids  \cite{Barabasi02}, although since we have restricted the number of neighbor-neighbor contacts this is not a perfect analogy.

\begin{figure}
\centering
\includegraphics[width=3.2in]{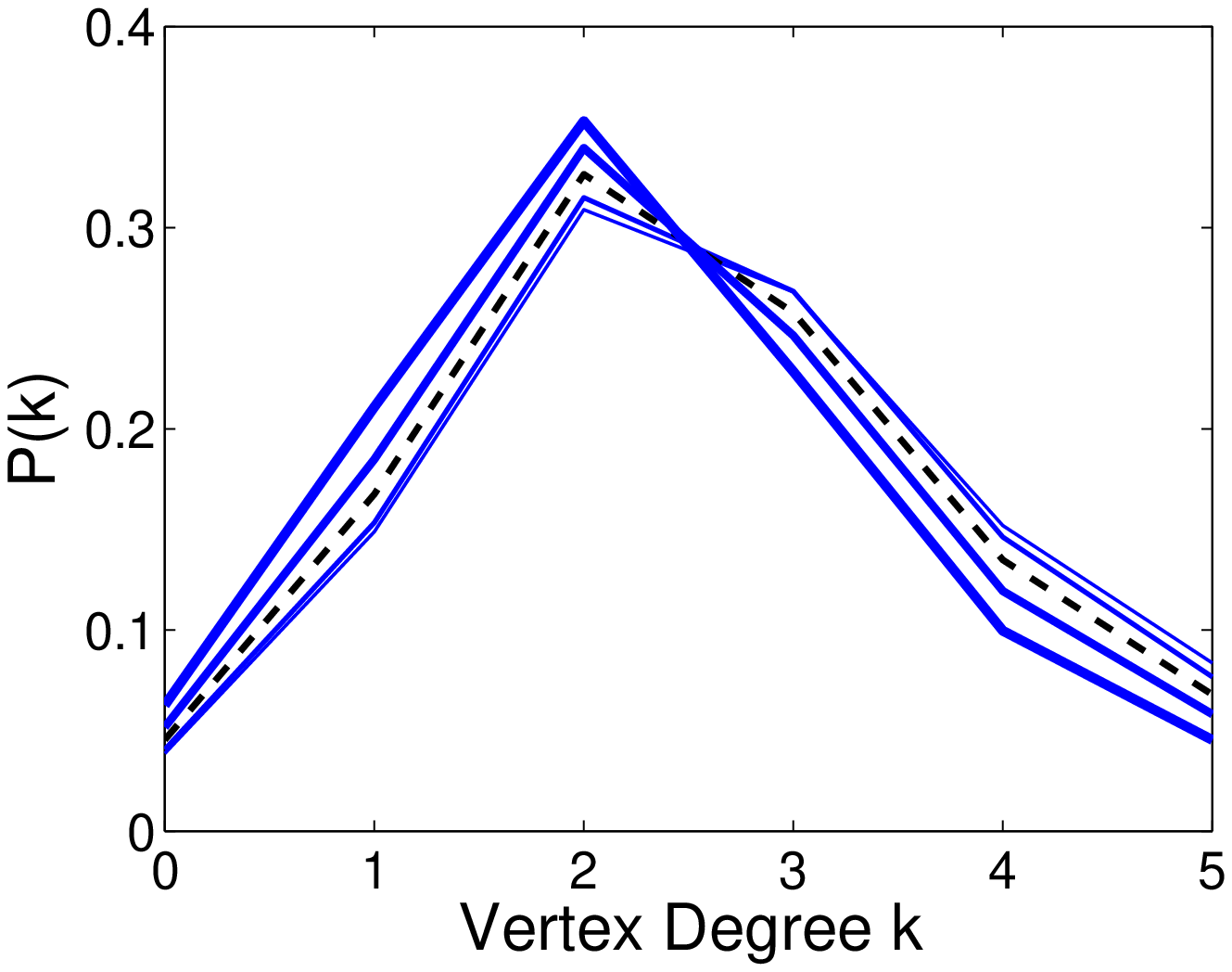}
\includegraphics[width=3.2in]{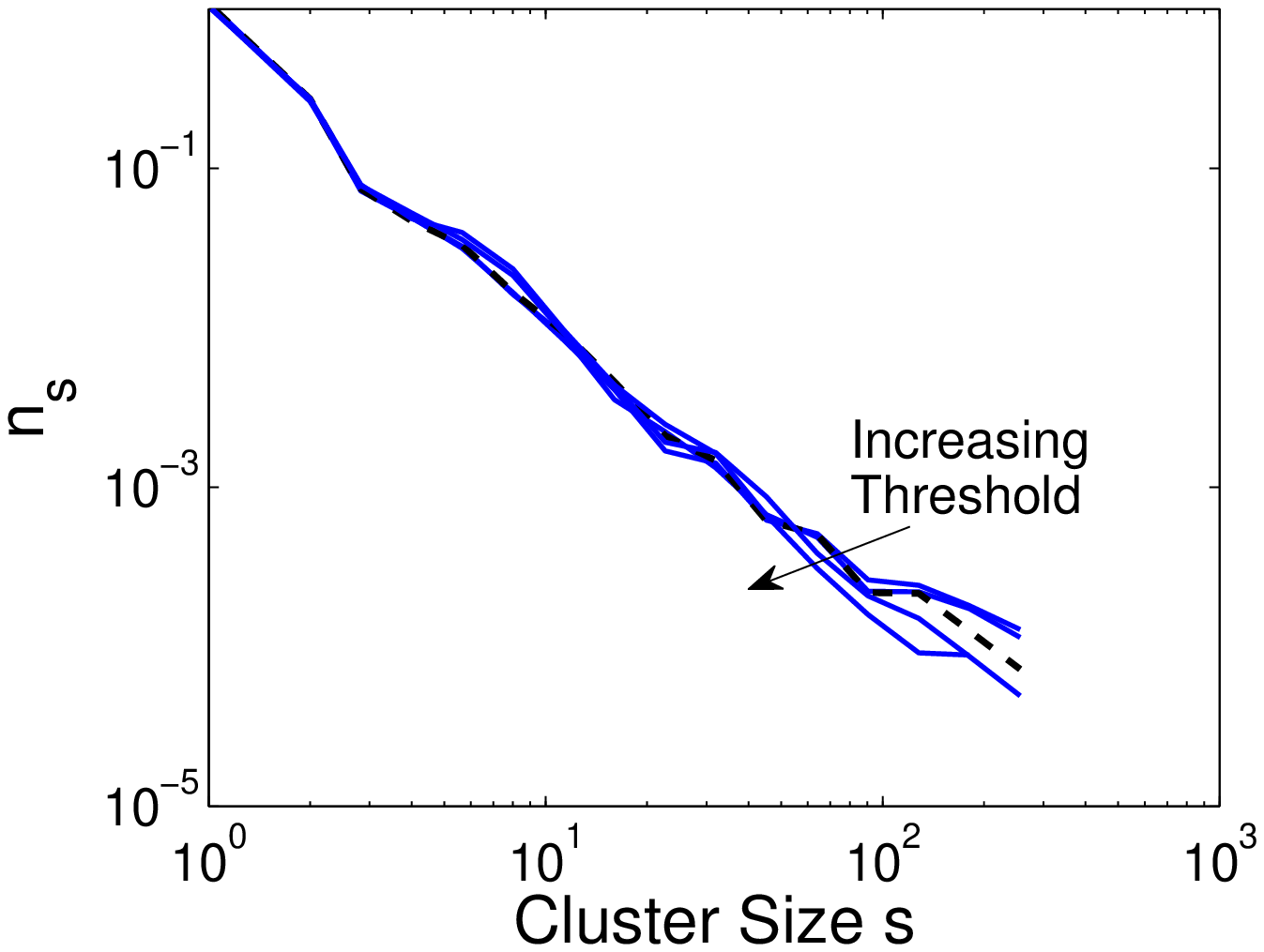}
\caption[Distributions for experimental force networks as a function of disk-finding threshold]{
{\bf Top} Distribution of vertex degree for experimentally obtained force
networks as a function of acceptability threshold for convolution with
our fringe kernel.  The dashed
line indicates the distribution for the acceptability threshold of $0.5$
used for the analysis in this paper.  Lowering the threshold from
$0.6$ (thickest line) to
$0.4$ (thinnest line) in increments of $0.05$
increases the number of vertices ``found"
and skews the distribution away from chain-like structures towards higher
vertex degree.  {\bf Bottom} Despite the change in vertex degree
distribution, the distribution of cluster size still indicates a power law with the same slope for all thresholds
considered.  The dashed line again indicates $0.5$.}
\label{thresholdresults}
\end{figure}
Finally, we examine the robustness of our results.  To obtain force network graphs from our experimental data  we convolved our images with a single fringe image kernel that does not necessarily match all arrangements of fringes.  We used an
acceptability threshold of $0.5$, on a scale where $1.0$ is a perfect
match and $0.0$ is the opposite of the kernel.  As indicated by Fig.
\ref{thresholdresults}, varying the acceptability threshold between $0.4$ and $0.6$ only slightly affects properties like vertex degree distribution without
significantly changing the cluster size power law.

\subsection{Comparison with Spatial Models}

\begin{figure}
\centering
\includegraphics[width=1.06in]{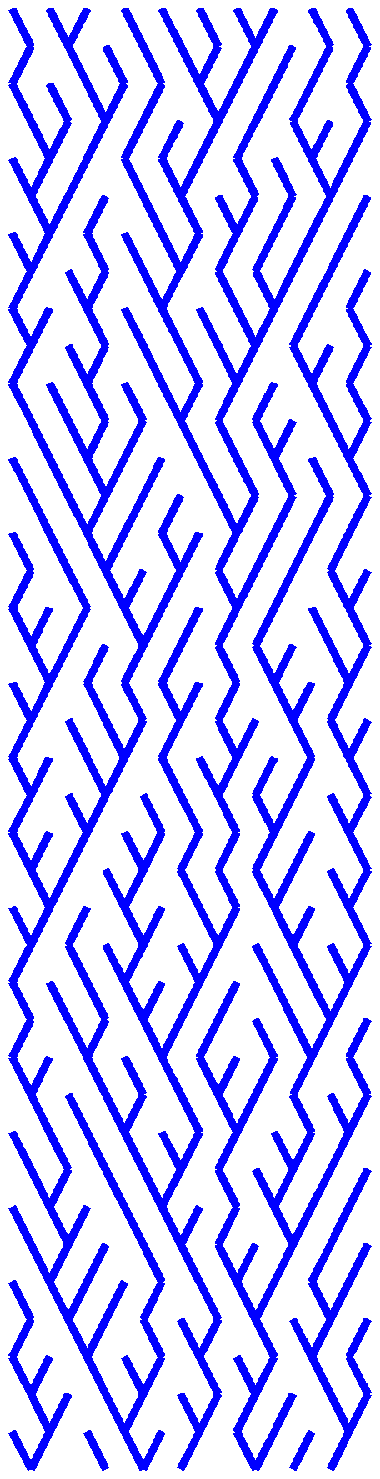}
\includegraphics[width=0.96in]{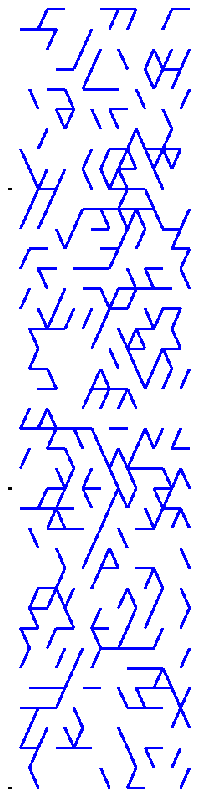}
\includegraphics[width=0.86in]{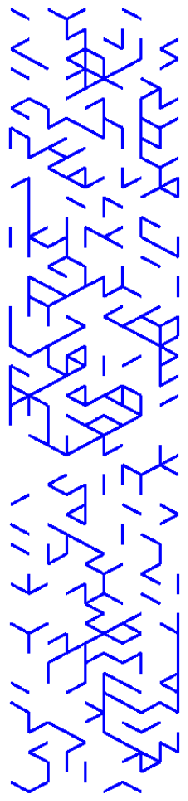}
\caption[Depiction of networks generated for geometric $q$-model and bond percolation on two different lattices]{From left to right: {\bf 1} An instance of the q-Model in the critical case. {\bf 2} A random network ($p = 0.25$) composed of edges from a triangular lattice oriented horizontally. {\bf 3} A random network ($p = 0.25$) composed of edges from a triangular lattice oriented vertically.}
\label{models}
\end{figure}
Having characterized actual force networks, it is possible to make comparisons to simple models.  Using an ensemble of realizations of the critical $q$-Model on a $10\times40$ vertex triangular grid (chosen to reflect the aspect ratio and roughly the number of vertices in our experiments) as in Fig.~\ref{models}.1, it was possible to determine distribution functions for vertex degree, cluster size, upward tree size and weights assigned by the $q$-Model.

In addition to the $q$-Model results, graphs were also constructed via bond percolation. Two regular triangular lattices were considered, one where in addition to downward-angled edges there were horizontal edges (see Fig.~\ref{models}.2), and one where there were vertical edges (Fig.~\ref{models}.3).  For each realization a subset of these edges were chosen with a probability $p$ to create a randomized graph.  With six possible edges per vertex, the triangular lattice is similar to the bi-disperse system though angular distributions are severely restricted.

A range of different $p$ values were examined for both triangular
orientations.  In addition to the $p = 0.2214$ that was determined
experimentally, we examined values of $p = 0.15$ and $p = 0.3$ above and
below the estimated $p$.  Finally, we examined $p = 0.5$, a value greater
than the percolation threshold, $p_c = 0.3473$, for a system-spanning
cluster in an infinite triangular lattice.

We generated $5000$ realizations each for the $q$-model and the two triangular lattice orientations.  For the bond percolation graphs both vertex degree and cluster size are
independent of whether the basis is oriented horizontally or vertically and so the joint ensemble of $10000$ realizations was analyzed.

\begin{figure}
\centering
\includegraphics[width=3.3in]{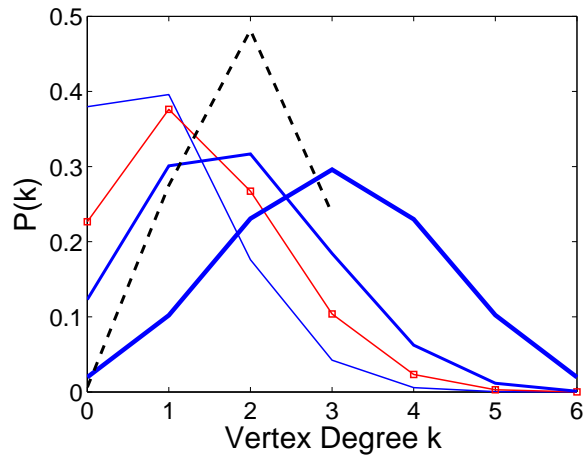}
\caption[Distribution of vertex degree for models]{The probability density function for
degree vertex for the critical $q$-Model and random triangular graphs.
The critical $q$-Model is indicated by a dashed line.  The bond percolation results for the experimentally obtained $p = 0.2214$ are indicated by $\Box$.  Other bond probabilities, $p = 0.15$, $0.3$ and $0.5$, are respectively indicated by lines of increasing thickness.  While the $q$-Model by design gives the expected peak at $k=2$ --- indicating chain-like structures --- since in the critical case of the $q$-Model $k \leq 3$, the best match to Fig.~\ref{forcedatak} appears to be $p = 3$.}
\label{modelkresults}
\end{figure}
The distributions of vertex degree for the critical
$q$-Model and triangular lattices with different bond probabilities are
shown in Fig.~\ref{modelkresults}.  Since the $q$-Model has to give chain
like structures, it is not surprising to see that $P(k)$ peaks at $k = 2$,
indicating that most elements are members of a single chain.  For a random
graph constructed with the bond probability $p = 0.2214$ calculated from
our experimental data, the distribution does not peak appropriately at $k
= 2$, but instead at $k=1$ indicating many pairs of bonds but no clusters.
Instead, the peak of the distribution for $p=0.3$ is most like the experimentally-observed distribution.  The distribution
for $p=0.3$ over-represents single bonds when compared with experiments, but the
distribution of larger $k$ values matches well.

Also plotted in Fig.~\ref{modelkresults} is the distribution for
$p=0.15$, below the measured $p$ and not agreeing with the experimentally
measured distributions. Finally, the distribution for $p = 0.5$, shows too
much skew toward large clusters.

\begin{figure}
\centering
\includegraphics[width=3.3in]{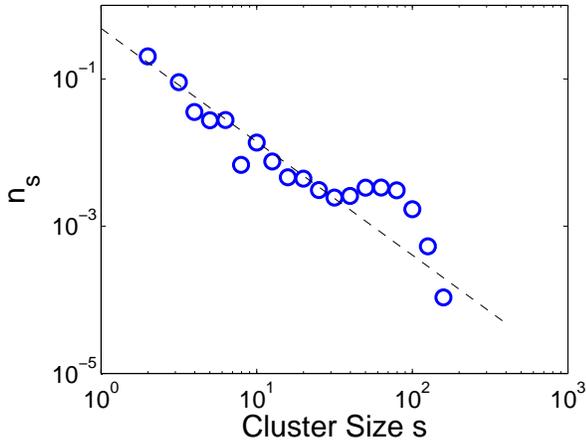}
\caption[Distribution of cluster size for $q$-model]{The distribution of number of clusters of size $s$ for the critical $q$-Model.  On a log-log plot the
distribution of smaller clusters is roughly linear (dashed line indicates slope $-1.53$ found by fitting to linear region), but there is a
peak at large cluster sizes since the critical
$q$-Model requires that each site has a downward neighbor.}
\label{qmodelsresults}
\end{figure}
\begin{figure}
\centering
\includegraphics[width=3.3in]{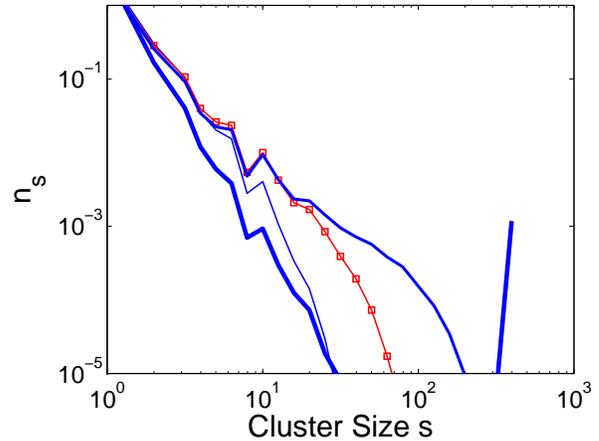}
\caption[Distribution of cluster size for bond percolation]{The distribution of the number of clusters of size $s$ for random triangular graphs.  The bond percolation results for the experimentally obtained $p = 0.2214$ are indicated by $\Box$.  Other bond probabilities, $p = 0.15$, $0.3$ and
$0.5$, are respectively indicated by lines of increasing thickness.
For all cases there is roughly linear behavior for small to moderate ($\sim100$) cluster sizes.  The
transition value to universal clusters is $p_c = 0.3473$ for an infinite triangular
lattice and as expected the slope varies
non-monotonically, decreasing as $p$ is
increased until abruptly dropping at $p = 0.5$ where a second peak at large
cluster size indicates that infinite clusters are forming.}
\label{modelsresults} 
\end{figure}
As indicated in Fig.~\ref{qmodelsresults}, the
$q$-Model gives a distribution (with logarithmic binning) of cluster size that falls linearly ($a_2 = -1.53$) on a
log-log plot until broadly peaking
at roughly a quarter of the
system size ($100$ vertices).  Because we have chosen the critical
case of the $q$-model to ensure long chains of contacts, it is not
surprising that we see more large clusters.  In contrast, the cluster
size distributions for graphs constructed from
triangular lattices shown in Fig.~\ref{modelsresults} are monotonic and roughly linearly, with decreasing slope as $p$ is increased until the critical threshold is crossed, after which the slope for small cluster sizes increases and a sharp peak at the size of the system appears, indicating infinite clusters (similar ``bi-modal"
distributions were observed previously bond percolation simulations
of much larger ($10^5$) systems \cite{Argyrakis03}).  The slope for $p = 0.3$ is found to be most similar to that of the experimentally observed networks: $a_2 = -1.77$ with $R^2 \approx 0.97$ compared to $a_2 = -1.6$ for experiment.

\begin{figure}
\centering
\includegraphics[width=3.3in]{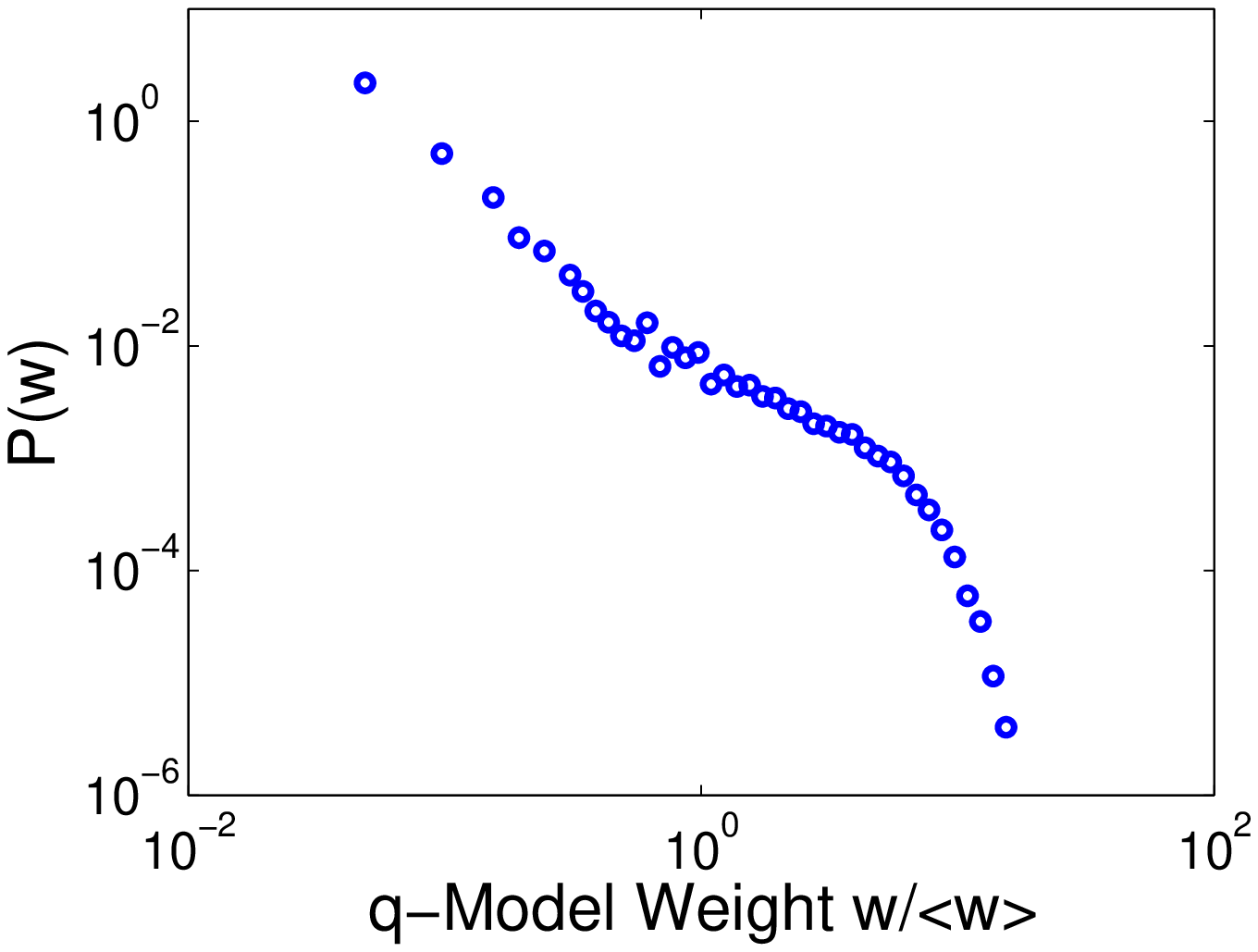}
\caption[Distribution of weights for $q$-model]{Probability
distribution of weight assigned by the $q$-Model for the critical $q$-Model graphs.  Because there are far more vertices with weights below the mean, the distribution of $q$-Model weights is
different from both the actual force and geometric $q$-Model weight distributions for
experimentally obtained force networks.}
\label{qmodelwresults}
\end{figure}
\begin{figure}
\centering
\includegraphics[width=3.3in]{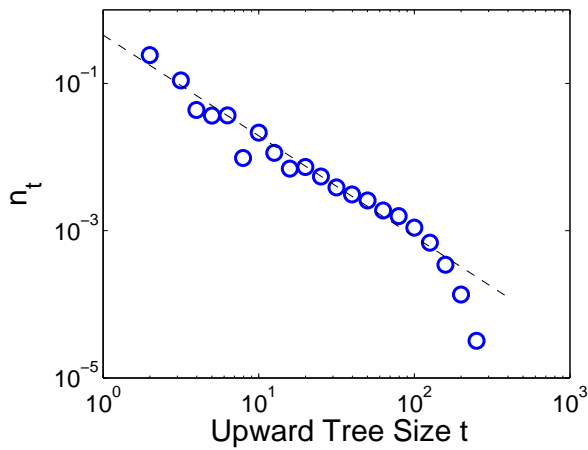}
\caption[Distribution of upward clusters for $q$-model]{Number
distribution of upward tree size for the critical $q$-Model.}
\label{qmodeltresults}
\end{figure}
Figure \ref{qmodelwresults} shows the distribution of ``weight" assigned via the $q$-model for the critical $q$-Model graphs.  The model distribution is more skewed toward large weights than the $q$-Model-like weight assigned to the experimentally obtained graphs.  This is unsurprising since, by construction, the critical $q$-Model is a connected graph.  The distribution of upward trees in Fig.~\ref{qmodeltresults} is roughly linear, with nearly the same slope ($a_2 = -1.52$) as cluster size.

\begin{figure}
\centering
\includegraphics[width=3.3in]{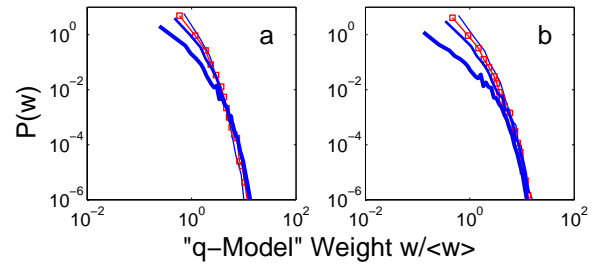}
\caption[Distribution of weights assigned by geometric $q$-model to bond percolation graphs]{Probability distribution of geometric $q$-Model weight for graphs assembled from horizontally- ({\bf
a}) and vertically- ({\bf b}) aligned triangular lattices.  The results for the experimentally obtained $p =
0.2214$ are indicated by $\Box$.  Other bond probabilities, $p = 0.15$, $0.3$ and
$0.5$, are respectively indicated by increasing line thickness.  The two alignments differ in the distribution of lower weights.  In both cases, only the largest, $p = 0.5$, approaches the distribution observed for experimentally-obtained
networks (Fig.~\ref{forcedatamodels}).}
\label{modelwhvresults}
\end{figure}
In Fig.~\ref{modelwhvresults}, we have the distribution of weights
assigned by the geometric $q$-Model.  None of the
models showed the experimentally observed peak for actual force found in Fig.~\ref{forcedataF}.  Graphs composed from edges in vertically-aligned
lattices --- where an additional downward edge is possible --- showed
slightly broader distributions than those that were composed from
horizontally-aligned triangular lattices.  In the
vertically-aligned
lattices, the vertices with weight below the mean are distributed more
broadly than in the horizontal case.  For both alignments, the
distribution for $p = 0.5$ was the most similar to
the distribution for graphs obtained from experimental data --- this
values is much larger
than the $p$ measured experimentally.  None of these distributions can be
said to be a good match for the force or $q$-Model distributions for experimentally
obtained graphs.

\begin{figure}
\centering
\includegraphics[width=3.3in]{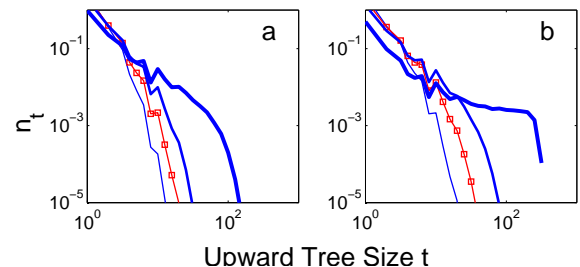}
\caption[Distribution of upward tree size for bond percolation graphs]{Distribution of number of upward trees of size $t$ for bond percolation graphs assembled from horizontally- ({\bf a})
and vertically- ({\bf b}) aligned triangular lattices.  The results for the experimentally obtained $p =
0.2214$ are indicated by $\Box$.  Other bond probabilities, $p = 0.15$, $0.3$ and
$0.5$, are respectively indicated by increasing line thickness.  Only when infinite clusters are present ($p = 0.5$)
do the distributions change significantly, in both cases giving a distribution that is linear over the same ranges as observed experimentally (Fig.~\ref{forcedatat}).}
\label{modelthvresults}
\end{figure}
As Fig.~\ref{modelthvresults} shows, the distributions of upward trees for graphs below the percolation threshold are similar.  For $p = 0.3$ and below, the distributions are well-described by the product of a power law and an exponential ($R^2 > 0.98$) however the coefficients are very different from the experimentally-observed distribution of upward trees.  For $p=0.5$ the distribution is qualitatively similar to the experimental distribution in Fig.~\ref{forcedatat}, but is not well-described by any of the distributions considered (power law, log-normal and power law-exponential).  Generally, when there are vertical edges the distribution of cluster sizes is broader.  This is reasonable since the addition of vertical edges makes large trees more likely.

\section{Conclusion}

We have characterized the graphs generated from analyzing experimentally-obtained force networks in granular material.  Assuming that there are an average of four contacts per grain, we calculate that the probability that
a contact carries a large force is $p = 0.2214$.  We observe evidence for a power law scaling of clusters of bonds with large force, possibly indicating a simply-characterized underlying geometry.

When we compare our experimental networks to bond percolation using random sets of edges from triangular lattices, we find that geometric properties of force networks can indeed  be matched depending upon the probability of the presence of a bond $p$.  We find that we cannot match all properties of the experimental networks using any one bond probability, including $p = 0.2214$.  However, the value of $p = 0.3$ worked well for both the vertex degree and cluster size distribution.

A slightly larger value of $p$ may be an artifact of under-sampling the number of minimally stressed particles that are hard to detect visually but do contribute stressed bonds.  It is possible that further refinement of the graph finding algorithm is needed to get the correct value of $p$ for actual force networks.

Because of obvious similarities, we have made comparisons with the $q$-Model and attempted to assign weight to, and determine ``force" distribution of, vertices in both $q$-Model and bond percolation graphs.  None of the methods studied provide good agreement with the experimentally-obtained force distribution.  This indicates that though graph theoretic approaches may in fact capture the geometry of force networks, the actual force distributions require more elaborate, vector analysis.

\section*{Acknowledgements}

I thank John Reif for suggesting studying the applications of graph theory to force networks; Xiaobai Sun and Nikos Pitsianis for computer science advising; Richard Palmer, Josh Socolar, Brian Tighe and Trush Majmudar for helpful conversations; and my Ph.D. advisor Robert Behringer.  This research was funded by National Science Foundation grants DMR-0137119 and DMS-0204677 and NASA grant NNC04GB08G.

\bibliography{granular}
\end{document}